\documentclass{seminarpaper}
\usepackage{url}
\title{\textsc{Biases in Data Science Lifecycle}}
\date{}
\author{
  \textsc{Dinh-an Ho}\\
  \texttt{dinh-an.ho@rwth-aachen.de}
  \and
  \textsc{Oya Beyan}\\
  \texttt{beyan@fit.fraunhofer.de}
}

\begin{document}

\maketitle

\begin{abstract}
In recent years, data science has become an indispensable part of our society. Over time, we have become reliant on this technology because of its opportunity to gain value and new insights from data in any field - business, socializing, research and society. At the same time, it raises questions about how justified we are in placing our trust in these technologies. There is a risk that such powers may lead to biased, inappropriate or unintended actions. Therefore, ethical considerations which might occur as the result of data science practices should be carefully considered and these potential problems should be identified during the data science lifecycle and mitigated if possible.  However, a typical data scientist has not enough knowledge for identifying these challenges and it is not always possible to include an ethics expert during data science production. The aim of this study is to provide a practical guideline to data scientists and increase their awareness. In this work, we reviewed different sources of biases and grouped them under different stages of the data science lifecycle. The work is still under progress. The aim of early publishing is to collect community feedback and improve the curated knowledge base for bias types and solutions. We would like to ask your feedback in one of the following ways:
\begin{center}
\begin{enumerate}
\item please participate in the survey \href{(https://forms.gle/tN75M9GMBgTj3rCH7)}{https://forms.gle/tN75M9GMBgTj3rCH7} or
\item leave your comment to the Google Docs \href{(https://bit.ly/35brLic)}{https://bit.ly/35brLic}. 
\end{enumerate}
\end{center}

\end{abstract}
\section{Introduction}
Data science has developed into an indispensable part of our society. We use this technology to collect information and extract knowledge from data. Almost all sectors use data science to achieve this. There are more and more new innovations, so that this works more and more efficiently. However, we have to think about ethical risks. Since human data is often involved, biases can occur. The ethical consequences that could result from data science-related practices should therefore be carefully considered and potential problems should be identified and, if possible, mitigated during the life cycle of data science. However, a typical data scientist does not have sufficient knowledge to identify these challenges, and it is not always possible to involve an ethical expert in data science production. The aim of this study is to provide a practical guideline to data scientists and increase their awareness. In this work we described different sources of biases in each stage of data science, provided some examples and gave references to the best practices. This work is conducted as part of master thesis  "An Ethics guideline for data scientist: developing an executable guideline for responsible data science" at the \textsc{RWTH Aachen University, Chair of Computer Science 5 Information Systems}. The main goal is to give data scientists an overview of the biases that can occur when programming and analyzing data,  and how these problems can be solved.The final outcome of the thesis will be a practical tool which data scientists can use in their daily work environment for accessing the related information and reporting their solutions for identified biases in their work. The thesis is still under progress.\\
\\
In this paper, we will present the outcome of the literature review. The next sections, first we will present our methodology, then provide the conceptual steps of a data life science cycle. And then present a collection of sources of biases in each phase with an example and with best practices for mitigating them. 
The findings in this document are still incomplete at this stage and will be completed in the progress of the master thesis.  The aim of early publishing is to collect community feedback and improve the curated knowledge base for bias types and solutions. 

\section{Research Methodology}
The primary purpose of this chapter is to figure out different types of biases that appear throughout the phases of the data science pipeline and map each bias with its unique phase. In order to do this multidisciplinary task, we explored 85 articles (cited properly) that belong to their different journals using Google Scholar, IEEE digital library, ACM digital library, Wiley online library, Semantic Scholar, Science direct, arXiv and so forth. In this chapter, we reviewed these articles and collected relevant information regarding the source of biases and mitigation methods. Then we mapped each source to the data science life cycle. Each phase of the data science pipeline is studied in terms of respective biases that can appear in that phase along with the Use-case examples and best practices. Data masters can utilize these best practices amid these biases.\\
\\
The input for the search is based on the data science phases, for example, one input was 'Data ingestion bias'. Analogously, the other biases were found. Examples were found based on the most read papers for the specific biases. In parallel, examples or best practices (here for example in the case of algorithmic bias) were searched by keywords like 'algorithmic bias example', 'algorithmic bias best practices', 'algorithmic bias mitigation'. For papers with keywords matching were collected and added to the search for papers.

\section{Phases in the Data Science Lifecycle}
In this research, we list all biases at each phase so that we have an overview of the biases that occur at that phase. There are several ways to define the phases of data science. Figure 1 shows one possibility which are as follows:
\begin{enumerate}
\item \textbf{Data Ingestion:} here the data are collected and imported by the data scientists from databases or self-produced data collections
\item \textbf{Data Scrubbing:} here the data is cleaned so that machines can understand and process it
\item \textbf{Data Visualization:} here significant patterns and trends are filtered by statistical methods
\item \textbf{Data Modeling:} here data models are predicted and forecasted using e.g. artificial intelligence algorithms
\item \textbf{Data Analysis:} here the results are interpreted, and knowledge is extracted.
\end{enumerate}

\begin{figure}
    \includegraphics[width=\columnwidth]{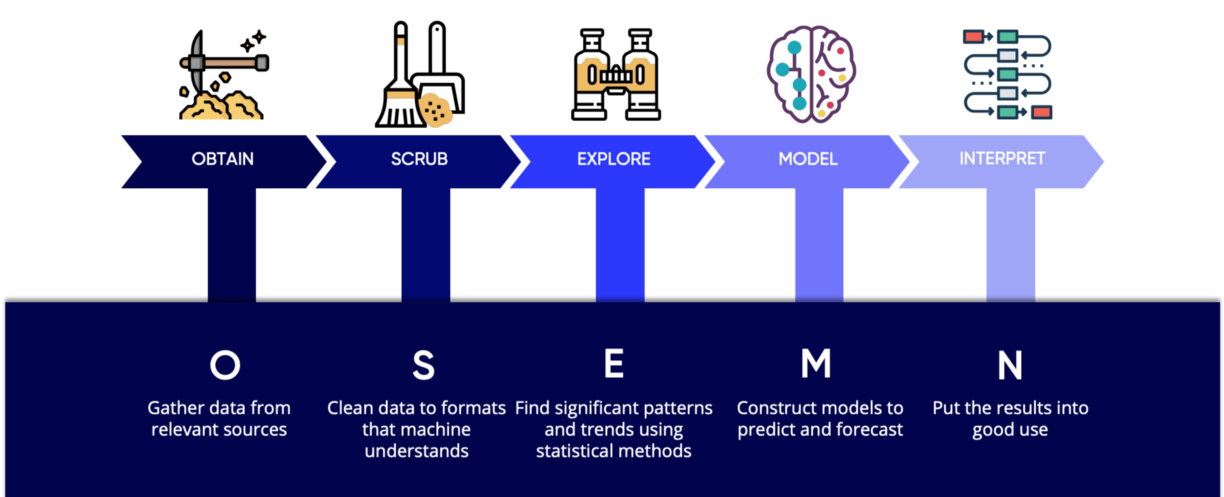}
    \caption{\label{fig:osemn2}This is the data science pipeline proposed by Dineva et al. \cite{osemn}}
\end{figure}

Another model by Suresh et al. \cite{P15} describes the following phases: world, population, dataset, training data, model, model output and real world implications. You can see that there are parallels between the two models.  The first model started directly from the data set and even included the data scrubbing phase, as this is an important part of data science analysis. However, this model has included the training data in this phase. The modeling and analysis phase is identical.  There are other well-known models, but we agree to use Figure \ref{fig:osemn2} as a lifecycle.

\section{Biases during each phase of Data Science Lifecycle}
In this section we iteratively go through all data science phases and indicate which biases may occur. All biases are divided into a short description, examples and best practices. An overview of all biases is shown in Figure \ref{fig:to3}. 
\begin{center}
\begin{table}[h]
\centering
\small
\begin{tabular}{l | l | l | l | l}
Data Ingestion & Data Scrubbing & Data Visualization & Data Modeling & Data Analysis \\
\hline
Data B. & Exclusion B. & Cognitive Biases: & Algorithmic B. & Deployment B.\\
Sampling B. & Data Enrichment B. & Framing Effect & Hot Hand Fallacy & Rescue B.\\
Measurement B.&&Availability B.&Bandwagon B.&Overfitting\\
Survey B.&&Overconfidence B.&Group Attribution B.&Underfitting\\
Seasonal B.&&Anchoring B.&Aggregation B.&\\
Survivorship B.&&Confirmation B.&&\\
Selection B.&&and Signal Error&&\\
Historical B.&&&&
\end{tabular}
\caption{\label{fig:to3}Overview of biases}
\label{tab:abc}
\end{table}
\end{center}
\subsection{Data Ingestion}
In this phase, the data are collected and imported by the data scientists from databases or self-produced data collections. The following describes the biases that can occur in this stage.
\subsubsection{Data Bias (Representation Bias)}
Data bias is a \textit{systematic distortion in the data that compromises its representativeness}. It is directly related to sampling that confirms whether the sample is representative of the larger population or not. It occurs during the data accumulation phase. Although capturing bias-free dataset is not feasible, data scientists can estimate the biasness in the data by comparing the sample with multiple samples having different contexts \cite{P1}.
\paragraph{Example:}In \cite{P2}, the writers claimed that machine learning algorithms discriminate against people on race, gender and ethnicity. They represent that the databanks of Adience and IJB-A primarily hold light-skinned subjects — 86.2 percent and 79.6$\%$ respectively, which can bias towards the underrepresented dark skin groups.
\paragraph{Best practice:}In \cite{P3}, novel data Representation bias Removal (REPAIR) technique is introduced to resolve the generalization issues present in training datasets by employing Repair-Algorithm. Proper labeling of the data just like the nutrition chart is another way to reduce data bias by task-oriented categorization of data \cite{P4}. Using supporting sheets for datasets can be valuable while lessening the data bias. Advanced data mining and proper data targeting are some other options for data recruiters to hire in order to obtain less discriminated data \cite{P5}.
\subsubsection{Selection Bias}
Selection bias occurs when \textit{wrong} contributors are selected or allowed to contribute, and therefore, the proper randomization is not achieved.
\paragraph{Example:}Selection bias occurs for example, if the residents of rural areas were selected to participate in a city transportation network related project. Due to the nature of VGI projects selection bias is one of the most important and influencing types of biases and also relatively hard to detect and treat \cite{P6}.
\paragraph{Best practice:}To correct this bias, it is especially important to ensure that selection bias is avoided when recruiting and retaining the sample population. Picking up subgroups randomly can also be beneficial to limit selection bias \cite{P7}.
\subsubsection{Sampling Bias}
Sampling bias is closely related to selection bias that can occur when all the members of data are not objectively representative of the target population about which conclusions are to be drawn. In addition, errors in the process of collecting samples generate sampling bias while errors in the subsequent processes cause selection bias \cite{P8}. Non-representative samples often lead to models that exhibit systematic errors. In biased sampling, the whole dataset is divided into two groups namely minority and majority groups. Hence the model may be trained according to the dominating and prejudicial behavior of the assessments. Therefore, proper selection of training data is the crucial part for data scientists as it is extremely challenging for them to compute the ground reality \cite{P9}.
\paragraph{Examples:}
\begin{enumerate}
\item Tainted training examples might wrongly instruct the machine to see features that actually predict success on the job as indicators of poor performance.
\item A classic example of a biased sample happens in politics, such as the famous 1936 opinion polling for the U.S. presidential elections carried out by the American Literary Digest magazine, which over-represented rich individuals and predicted the wrong outcome \cite{P10}.
\end{enumerate}
\paragraph{Best practice:}If some groups are known to be under-represented and the degree of under-representation can be quantified, then sample weights can correct the bias \cite{P11}. Simple random sampling, and Stratified random sampling are some valuable tricks to mitigate sampling bias \cite{P12}. Stratification involves the division of the whole population into different subgroups, for instance, measuring the similar attributes of multiple subgroups under same conditions. Hence, this approach offers in-depth inspection of the relations among the groups and highly précised scores as variability is low in homogenous groups \cite{P13}.
\subsubsection{Measurement Bias (Systematic Errors)}
Measurement bias arises when the data analyst tries to get desired results by selecting, operating, and measuring a particular feature \cite{P5}. This may cause us to skip some important factors or create group or noise in the process that would lead to disaster. It relates with the data bias to some extent but the main difference is that data bias is due to inherent bias in the data that gives biased outcome due to non-standardize data while, on the other hand, includes the addition of unnecessary data unconsciously or deliberately.
\paragraph{Example:}Measurement bias was spotted in the risk prediction tool COMPAS, in which the former arrests and family arrests were employed as proxy variables to measure level of crime. COMPAS predicted incorrectly based on dissimilar subgroups. As marginal groups are controlled and policed more often, thus they have a higher rate of arrests \cite{P15}.
\paragraph{Best practice:}Systematic errors cannot be avoided simply by collecting more data, but having multiple measuring devices (or observers of instruments), and data specialists to compare the output of these devices \cite{P11}.
\subsubsection{Survey Bias (Self-Reporting Bias)}
As per the name, survey bias occurs when researchers receive prejudiced, inconsistent and tailored feedback or no feedback to the interviews, surveys and questionnaires from the respondents. The main behind all that is the presence of secret or sensitive topics of discussions concerning income, sex, drugs, race, violence etc. Consequently, self-reporting data can be influenced by two types of external bias: (1) Social approval (that can underestimate the original value); and (2) Recall Bias (mostly people recall and answer that might be erroneous) \cite{P11}. The concept of data linkage is very important to understand for the data analysts here. Data linkage involves in the data collection process when information about the same entity is gathered from two or more distinct sources. Probabilistic matching and Individual reference identifiers can be useful while joining two or more survey datasets \cite{P16}.
\subsubsection{Seasonal Bias}
Sometimes the available data is related to seasonal entities which simply means that the dataset exhibits seasonal growth patterns. Data interpreters when considering this kind of situational data (Time Series) for training the supervised models, they are said to be seasonally biased. Additionally, predictive models are gravely impacted by seasonality because of the dynamic fluctuations present in the records.
\paragraph{Example:}The Indian financial year ends on 31st March. It is a high time for Indian insurance industry because people tend to buy more insurance items to claim the rebates at that time. An analysis of past 10 years of insurance business data shows that 25-30$\%$ of business of insurance industry in India come in the month of March. Similarly, there is a surge in sales of consumer goods in the UK and US leading up to Christmas \cite{P18}. Data analyzers have to keep track of trends that data has been facing since previous years to get optimized results from their models.
\paragraph{Best practice:}Having in-depth knowledge about the seasonal trends of the targeted industry is essential to avoid seasonal bias. Targeted industry is used as a generalized term, that includes the industry under examination, regardless of which industry it is (examples are given above). Data specialists can compare the values of peak time with normal day value and they should measure only what they need. Reviewing the historical trend to predict future patterns can also be the good approach for seasonal adjustments \cite{P19}.
\subsubsection{Survivorship Bias}
Survivorship bias occurs when only certain successful subsets of a group are considered while the failures are dropped out of observation. This type of dataset selection skews the average output upward showing fake performance \cite{P20}. Data scientists when they try to make sense out of incomplete data, they fell prey to the survivorship bias.
\paragraph{Example:}During World War II, researchers from the non-profit research group the Center for Naval Analyses were tasked with a problem. They needed to reinforce the military’s fighter planes at their weakest spots. To accomplish this, they turned to data. They examined every plane that came back from a combat mission and made note of where bullets had hit the aircraft. Based on that information, they recommended that the planes be reinforced at those precise spots. The problem, of course, was that they only looked at the planes that returned and not at the planes that didn’t. Of course, data from the planes that had been shot down would almost certainly have been much more useful in determining where fatal damage to a plane was likely to have occurred, as those were the ones that suffered catastrophic damage \cite{P21}.
\paragraph{Best practice:}Data scientists may alleviate survivorship bias in backtest with survivorship bias free datasets and/or more recent data. The former one includes information of delisted equities during the test period while it is likely that fewer stocks are delisting in a more recent, shorter time period \cite{P22}.
\subsubsection{Historical Bias}
The pre-existing bias arises due to social and technical disagreements in the world as it is; and it seeps into the data even after selecting features and collecting samples perfectly \cite{P15}.
\paragraph{Example:}In 2018, an image search for women CEOs showed fewer results as there were only 5 percent women in Fortune 500 CEOs. Even the output was the entire reality, algorithms should consider or avoid these inherent discrimination is the hot topic for data scientists \cite{P15}.
\paragraph{Best practice:}Recognizing historical bias requires a retrospective understanding of the application and data generation process over time. Historical bias is often concerned with demonstrating the representational harms (such as reinforcing a stereotype) against a distinct group \cite{P15}.
\subsection{Data Scrubbing}
This phase is also known as data cleaning phase. In this phase, the data is cleaned so that machines can understand and process it. The two biases that can occur in this stage are described below.
\subsubsection{Exclusion Bias}
Data cleansing is an essential phase of the data science lifecycle that comes after data collection. In ethical perspective, removal of corrupt or unethical data involving both upper and lower extremes and exceptions is crucially important. For instance, outliers (values that deviate from the pattern) and duplications, from big raw data to make it less redundant, more consistent and reliable for model training. Since excluding un-actionable and duplicate insights is an important part while cleaning noise from the data, experts get biased while they get into it. Exclusion bias occurs when data handlers do not identify and remove the undesired chunks of data that should be removed in order to make data ethical and to maintain the accuracy of the results \cite{P23}.
\paragraph{Examples:}
\begin{enumerate}
\item We do customer profiling and find out that the average annual income of customers is 0.8 million dollar. But, there are two customers having annual income of 4 dollar and 4.2 million dollar. These two observations will be seen as outliers. Exclusion bias will occur if data managers do not exclude these two customers as their annual income is much higher than the rest of the population \cite{P24}. Exclusion bias can also arise when some important chunk is deleted from the data source while refining the data.
\item Exclusion of subjects who have recently migrated into the study area (this may occur when newcomers are not available in a register used to identify the source population). Excluding subjects who move out of the study area during follow-up is rather equivalent to dropout or non-response, a selection bias in that it rather affects the internal validity of the study \cite{P25}.
\end{enumerate}
\paragraph{Best practice:}To avoid this bias, file manipulators and data scientists should have intimate knowledge of data attributes, database sources as well as the data collection process in order to verify what to exclude or not (ethical exclusion). Secondly, extreme values from the data can be precisely detected through various techniques including k-Nearest Neighbor technique, ARIMA methodology, Regression analysis \cite{P26} or an Angel-based Outlier Detection \cite{P27}.
\subsubsection{Data Enrichment Bias}
After excluding outliers, data should be structured properly via managing the abnormal segments and missing segments. Organizing the available data means eliminating typos or grammatical mistakes from the data, and handling the missing data that is because of incomplete responses from the participants (Non-response bias). Data enrichment bias occurs due to typing mistakes of data entry operators or when they misinterpret the context of data and add the wrong input (extreme input) to fulfill empty fields \cite{P28}. Data cleaning though is the time consuming task but it can ultimately improve the decision making process.
\paragraph{Example:}Imagine a data entry operator mistakenly type \textit{Hspital} instead of \textit{Hospital} or evaluates the student pass having less than 50$\%$ marks while others are fail in the list having same percentage.
\paragraph{Best practice:}The serendipitous search in AI algorithms has enough potential to mitigate data-enrichment bias by exploring the unexplored parts through different ranking parameters \cite{P29}. A quality assurance committee having diverse experience of several disciplines should be formalized to review the data sources repetitively from the lens of morality to ensure fairness and reduce data discrimination \cite{P29}. Using techniques like Kernel-based local outlier factor (LOF) to identify the incorrect data can be helpful \cite{P28}. Furthermore, data dealers can create a separate category for records having missing values or flag them with 0 (if numeric) to make the algorithm aware of it \cite{P30}.
\subsection{Data Visualization}
Here significant patterns and trends are filtered by statistical methods. Cognitive bias can occur in the data visualization phase. This bias is divided into different types.
\subsubsection{Cognitive bias}
Studies have shown that visual transformations of data actually affect the data that impacts both decision making and the results. Tools that present data into the visual formats always try to make visualization easy for the auditors and by doing so, they may alter the original pattern of data \cite{P31}. Such improperly created graphs can trigger cognitive biases for the viewers \cite{P32}. Five types of cognitive biases are discussed in \cite{P33}.
\begin{enumerate}
\item Data Visualization and Framing Effect: 
Individuals respond to a particular problem in different ways depending upon how the problem is framed to them, a bias is called framing effect \cite{P34}. Data visualization tools often prioritize the data in the most comprehensible way. Meanwhile, they mostly alter the original sequence of data and trigger the framing bias for the auditor to cope with.
\begin{itemize}
\item[] \textbf{Example}: Graph highlights that 30 percent (70 percent) of a client's usual trade credits from suppliers are denied (awarded) might impact auditors' assessment of the probability and severity of their audit client's financial difficulties, which is one of the key conditions of an entity's ability to continue as a going concern (AU 341.06 An Entity's Ability to Continue as Going Concern, Consideration of Conditions and Events). Due to the framing effect, auditors who receive or process the negatively framed information (credit denial) are more likely to have substantial doubt about the client's ability to continue as a going concern. Therefore, improperly designed visualizations can trigger and/or aggravate framing effects \cite{P33}.
\item[] \textbf{Best practice}: Data analysts should use visualization tools strategically, let say, use an effective tool at the early stage of lifecycle \cite{P35}.
\end{itemize}
\item Data Visualization and Availability Bias: Availability bias relates to the survivorship bias expressing the tendency to use the already available information and consider such information more relevant than evidence that is hard to attain. Data visualization directly enhances the vividness and evaluability of data that influence availability bias but on the other hand, it compromises the overall quality of making effective decisions \cite{P31}.
\begin{itemize}
\item[] \textbf{Example}: After reading an article about lottery winners, you start to overestimate your own likelihood of winning the jackpot. You start spending more money than you should each week on lottery tickets \cite{P36}, \cite{P37}.
\item[] \textbf{Best practice}: People should spend proper time and effort contemplating other options, to properly weight them in terms of how well they meet the objective, or to consider the reliability, validity, certainty and accuracy of information \cite{P38}.
\end{itemize}
\item Data visualization and Overconfidence Bias: Overconfidence bias refers to an analyst's tendency to \textit{overestimate their own ability to perform tasks or to make accurate diagnoses or other judgments and decisions} \cite{P39}. Generally, the decision makers feel more confident in the graphical visualization of the data as compared to the textual format. Overconfidence leads to less cautious behavior that can be dangerous while making sensitive decisions.
\begin{itemize}
\item \textbf{Example}: In a survey conducted of 300 fund managers, asking if they believe in their managerial abilities with options average, above average and below average. Figures show that 74 percent believed that they were above average and of the remaining 26 percent thought they were average. While no one thought that he/she was below average. So, it is clear that these findings are statistically impossible or manipulated which is not suitable for data modeling \cite{P40}.
\item \textbf{Best practice}: Overconfidence can lead to overestimation and over-precision that is intolerable in statistical analysis. Therefore, one can channelize his/her overconfidence by creating scientific a mindset, by challenging his viewpoints, by listening to the criticism and by constant learning attitude \cite{P41}.
\end{itemize}
\item Data visualization and Anchoring Bias: Anchoring bias refers to the situation in which individuals rely too much on the initial piece of information offered and make future decisions by using this information. While visualizing data, anchoring bias may disturb the future interpretations and evaluations of insights coming from the same data based on preliminary evidence \cite{P33}.
\begin{itemize}
\item \textbf{Example}: E-commerce stores take benefits by using anchoring techniques by showing costly things first. Seeing 500 dollar shirts first and 60 dollar shirts at the second place, one will be prone to see the second shirt cheap.
\item \textbf{Best practice}: Critical thinking can be beneficial in avoiding it. One can study his/her own anchoring behavior and analyze its prospects. Making future decisions based on historical patterns can also limit anchoring bias. Asking your colleague for the review is not a bad choice \cite{P42}.
\end{itemize}
\item Data Visualization, Confirmation Bias and Signal Error: One can be the victim of both confirmation bias and signal error when there is a huge amount of data in hand. Signal error occurs when data analysts just overlook the major gaps between the data that make it inconsistent or unreliable. On the other hand, confirmation bias is the situation in which model builders unconsciously process the subset of data visualization that confirm their prior feelings and viewpoints \cite{P43} \cite{P33} . In addition, a trainer may actually keep training a model until it produces a result that aligns with their original hypothesis; this is called experimenter's bias. All of these biases can impede while making decisions.
\begin{itemize}
\item \textbf{Example}: Peter O. Gray \cite{P44} in his book presents an example of confirmation bias in the doctor's diagnosis. He explained that a doctor forecasts the disease after asking some queries from the patient and looks for the evidence that tends to confirm his/her diagnosis while overseeing the sign that inclines to defeat his analysis. Same is the case with data scientists who mostly tend to ignore the data that contradicts their hypothesis that ultimately have negative impacts on the process. Therefore, a data scientist should know all his biases and think scientifically to avoid such blunders.
\item \textbf{Best practice}: Confirmation Bias can be countered by continuously challenging your thoughts, by finding alternative sources of information and testing it \cite{P45}.
\end{itemize}
\end{enumerate}
\subsection{Data Modeling}
In this phase, data models are predicted and forecasted using e.g. artificial intelligence algorithms. 
\subsubsection{Algorithmic Bias}
Studies have shown the probability of unfairness in data is much greater than that of algorithms. More precisely, datasets are previously discriminated before passing through the algorithms that exhibit biased decisional pictures \cite{P46}. Machine learning algorithms based on AI, are commonly used while training the models in a supervised learning framework. Fairness is an increasingly important concern as autonomous models are used to support decision making in high-stakes applications such as mortgage lending, hiring, and prison sentencing \cite{P47}. To understand the responsibility of model failure, understanding the accountability matrix for algorithms is essential. Algorithmic bias is when an algorithm does not neutrally extract or transform the data. Scholars are trying hard to figure out the ways of mitigating the algorithmic biases present in Google searches, Facebook feeds, or in YouTube recommendations \cite{P48}.
\paragraph{Example:}Online retailer Amazon, whose global workforce is 60 percent male and where men hold 74 percent of the company's managerial positions, recently discontinued use of a recruiting algorithm after discovering gender bias. The data that engineers used to create the algorithm were derived from the resumes submitted to Amazon over a 10-year period, which were predominantly from white males. The algorithm was taught to recognize word patterns in the resumes, rather than relevant skill sets, and these data were benchmarked against the company's predominantly male engineering department to determine an applicant's fit. As a result, the AI software penalized any resume that contained the word \textit{women's} in the text and downgraded the resumes of women who attended women's colleges, resulting in gender bias \cite{P49}.
\\
\paragraph{Source of Algorithmic Bias\cite{P48}}
\begin{enumerate}
\item Biased training data can be the source of algorithmic bias.
\item Algorithms can be biased via differential use of information (using morally irrelevant categories to make morally relevant and sensitive judgments).
\item During the data processing, the algorithm can itself be biased, called "Algorithm Processing Bias". The most obvious instance of algorithmic processing bias is the use of a statistically biased estimator in the algorithm for better future predictions. So, this bias mostly occurs due to deliberate choice.
\item Algorithmic bias can also occur when the specific model is employed outside of its context, commonly known as Transfer Context Bias (for instance, using an autonomous system worldwide that was designed to be used in United States). It is basically the user bias but labeled as the \textit{algorithmic bias}.
\item Sometimes the information that an algorithm produces mismatch with the information that user expects, is known as Interpretation Bias. In the manual systems, misjudging the results of algorithms is actually the user bias but also notorious as \textit{algorithmic bias}. In autonomous systems, biased judgments about causal structure or strength (i.e., that deviate from the actual causal structure in the world) can easily be misused in biased ways by autonomous systems.
\item Algorithmic bias can occur when algorithms make decisions based on race, usually called racial bias. It may be due to the unintentional or open inclusion of racial characteristics by the developer in the databank or may be due to historical bias in data. Advanced health-care systems rely on commercial prediction algorithms to identify and help patients with complex health needs, therefore, there are high risks attached with the biased predictions. A clear example of racial disparity is that African American patients are considerably sicker than white patients, as evidenced by signs of uncontrolled illnesses \cite{P50}.
\end{enumerate}
\paragraph{Best practice:}AIF360 is the first system to bring together in one open source toolkit: bias metrics, bias mitigation algorithms, bias metric explanations, and industrial usability. By integrating these aspects, AIF360 can enable stronger collaboration between AI fairness researchers and practitioners, helping to translate our collective research results to practicing data scientists, data engineers, and developers deploying solutions in a variety of industries \cite{P47}. The fairness metrics and the algorithms the tool is using is shown \href{https://AIF360.readthedocs.io/en/latest/modules/metrics.html}{here} and \href{https://AIF360.readthedocs.io/en/latest/modules/algorithms.html}{here}. The algorithms used by AIF360 are the following:
\begin{itemize}
\item[] \underline{Pre-processing:}
\begin{enumerate}
\item \textbf{Disparate impact remover} is a preprocessing technique that edits feature values increase group fairness while preserving rank-ordering within groups \cite{P51}.
\item \textbf{Learning fair representations} is a pre-processing technique that finds a latent representation which encodes the data well but obfuscates information about protected attributes \cite{P52}.
\item \textbf{Optimized preprocessing} is a preprocessing technique that learns a probabilistic transformation that edits the features and labels in the data with group fairness, individual distortion, and data fidelity constraints and objectives \cite{P53}.
\item \textbf{Reweighing} is a preprocessing technique that weights the examples in each (group, label) combination differently to ensure fairness before classification \cite{P54}. 
\end{enumerate}
\item[] \underline{Post-processing:}
\begin{enumerate}
\item \textbf{Adversarial debiasing} is an in-processing technique that learns a classifier to maximize prediction accuracy and simultaneously reduce an adversary’s ability to determine the protected attribute from the predictions \cite{P55}.
\item \textbf{GerryFair Model} is an algorithm for learning classifiers that are fair with respect to rich subgroups \cite{P56}.
\item The \textbf{meta algorithm} here takes the fairness metric as part of the input and returns a classifier optimized w.r.t that fairness metric \cite{P57}.
\item \textbf{Prejudice remover} is an in-processing technique that adds a discrimination-aware regularization term to the learning objective \cite{P58}.
\end{enumerate}
\item[] \underline{Post-processing}
\begin{enumerate}
\item \textbf{Calibrated equalized odds postprocessing} is a post-processing technique that optimizes over calibrated classifier score outputs to find probabilities with which to change output labels with an equalized odds objective \cite{P59} .
\item \textbf{Equalized odds postprocessing} is a post-processing technique that solves a linear program to find probabilities with which to change output labels to optimize equalized odds \cite{P60}.
\item \textbf{Reject option classification} is a postprocessing technique that gives favorable outcomes to unprivileged groups and unfavorable outcomes to privileged groups in a confidence band around the decision boundary with the highest uncertainty \cite{P61}.
\end{enumerate}
\end{itemize}
There are many metrics that measure individual and group fairness. 
\begin{enumerate}
\item \textbf{Statistical Parity Difference:} This is the difference in the rate of positive results that the unprivileged group receives compared to the privileged group.
\item \textbf{Equal Opportunity Difference:} The difference in truly positive rates between unprivileged and privileged groups.
\item \textbf{Average Odds Difference:} The average difference between the false positive and true positive rate between unprivileged and privileged groups.
\item \textbf{Disparate Impact:} The ratio of rate of favorable outcome for the unprivileged group to that of the privileged group.
\item \textbf{Theil Index:} Measures the inequality in benefit allocation for individuals.
\item \textbf{Euclidean Distance:} The average Euclidean distance between the samples from two datasets.
\item \textbf{Mahalanobis Distance:} The average Mahalanobis distance between the samples from two datasets.
\item \textbf{Manhattan Distance:} The average Manhattan distance between the samples and two datasets.
\end{enumerate}
\paragraph{FAIRECSYS}It is an algorithm that mitigates algorithmic bias by post-processing the recommendation matrix with minimum impact on the utility of recommendations provided to the end-users \cite{P62}. By giving people control of their digital footprints can be helpful to reduce data bias. Algorithm transparency is another way to address the issue of algorithmic bias. Policy makers should design and implement discrimination-free laws to counter the lag in proper decision making that is, due to racial bias \cite{P63}.
\subsubsection{Hot Hand Fallacy}
In the data science lifecycle, this bias appears when data experts use a particular model repeatedly for all the problems based on its historical performance without testing other suitable models. In this phenomenon, a person who got best results recently is supposed to have greater chances of success in future \cite{P64}.
\paragraph{Example:}When we throw a coin 20 times, then there is a 50 percent chance of getting four heads in a row, a 25 percent chance of five in a row and a 10 percent chance of a run of six. However, if you give this sequence to most individuals they will consider that these were patterns in the data and not at all random. This explains the \textit{hot hand} fallacy in which we think we are on a winning streak – in whatever that may be – from cards to basketball to football. In each of these areas where the data is random but happens to include a sequence, we massively over-interpret the importance of this pattern \cite{P65}.
\paragraph{Best practice:}Data scientists should think systematically and treat each problem independently as per its requirements. If we again examine the coin toss example, just because you threw tails three times and won, it does not mean the fourth toss will also result in a win. Therefore, one should segregate each problem and try to make logical conclusions for the choices \cite{P66}.
\subsubsection{Bandwagon Effect}
The bandwagon effect is the type of cognitive bias that refers to the individual tendency to follow the behavior of the mass \cite{P67}. In model building, this effect appears when we are derived by the impulse to adopt a specific methodology just because it previously been adopted by others. Hence, the data scientists blindly select the model without any evaluation as human brains love shortcuts.
\paragraph{Example:}Tonsillectomy is cited as a recent example of medical bandwagons. Although the practice is said to be beneficial in some specific cases, scientific support for the universal use it saw was lacking. Doctors were drawn to tonsillectomy not on the basis of its effectiveness, but because they saw it was widely used \cite{P68}.
\paragraph{Best practice:}Despite Bandwagon effect can help to adopt healthy behavior, data analysts should think twice before jumping into it. One must re-think either he/she is rational or influenced by the environment/group while making influential judgments. One must evaluate the algorithms before rushing towards them without knowing about constraints \cite{P69}.
\subsubsection{Group Attribution Bias}
It is to believe that a person's traits always follow the ideologies of a group to which he/she belong and the decisions of the group manifest the beliefs it is every member \cite{P70}. This bias consist of two categories:
\begin{itemize}
\item In-group bias: A preference for members of a group to which you also belong, or for characteristics that you also share.
\begin{itemize}
\item[] \textbf{Example:} Two engineers training a résumé-screening model for software developers are predisposed to believe that applicants who attended the same computer-science academy as they both did are more qualified for the role \cite{P70}.
\end{itemize}
\item Out-group homogeneity bias: A tendency to stereotype individual members of a group to which you do not belong, or to see their characteristics as more uniform.
\begin{itemize}
\item[] \textbf{Example:} Two engineers training a résumé-screening model for software developers are predisposed to believe that all applicants who did not attend a computer-science academy do not have sufficient expertise for the role \cite{P70}.
\item[] \textbf{Best practice:} In order to avoid group attribution biases, data scientists should not behave judgmentally rather they should analyze the situation and efficiently respond to the situation. Emotional and Cultural intelligence is another skill that can be handy in mitigating fundamental attribution errors while scheming a model. Self-analysis is one of the versatile techniques to avoid severe favoritism \cite{P71}.
\end{itemize}
\end{itemize}
\subsubsection{Aggregation Bias}
Aggregation bias occurs while model creation, when one framework is used for the groups having distinct conditional distributions, $P(Y/X)$. In many applications, the concerned population is heterogeneous, and a single model does not fit all the subgroups. Aggregation bias can lead to a model that is suitable for the dominant population, or a model that does not fit any group at all (if combined with representation bias) \cite{P15}.
\paragraph{Example:}Diabetes patients have known differences in associated complications across ethnicities \cite{P72}. Studies have also suggested that HbA1c levels (widely used to diagnose and monitor diabetes) differ in complex ways across ethnicities and genders \cite{P73}. Because these factors have different meanings and importance within different subpopulations, a single model to predict complications is unlikely to be best-suited for any group in the population even if they are equally represented in the training data \cite{P15}.
\paragraph{Best practice:}Coupled learning techniques, for instance, multitask learning and Fair representation learning approach such that, space projection of data can be useful while countering aggregation bias \cite{P74}, \cite{P75}.
\subsubsection{Evaluation Bias}
Evaluation bias happens during model iteration and evaluation when the training or benchmark data do not represent the targeted population. Evaluation bias can also arise from the use of performance metrics that are not appropriate for the way in which the model will be used. This can be intensified by the use of inappropriate metrics that are arrayed to report performance boost \cite{P15}.
\paragraph{Example:}In Buolamwini and Gebru \cite{P2}, the example discussed under data bias in the context of the structure of face recognition refers to the drastically inferior performance of commercially used face analysis algorithms. Looking at some common facial analysis benchmark datasets, it becomes apparent why such algorithms were considered appropriate for use - 7.4 percent and 4.4 percent of the images in benchmark datasets such as Adience and IJB-A are of dark-skinned female faces. Algorithms that underperform on this slice of the population therefore suffer quite little in their evaluation performance on these benchmarks. The algorithms' underperformance was likely caused by representation bias in the training data, but the benchmarks failed to discover and penalize this. Since this study, other algorithms have been benchmarked on more balanced face datasets, changing the development process to encourage models that perform well across groups \cite{P76}.
\paragraph{Best practice:}To mitigate evaluation bias an approach namely Subgroup Evaluation can be used to comprehend the group matrices clearly by comparing them. Multiple metrics and confidence intervals is another useful technique in choosing relevant metrics for modeling \cite{P2}, \cite{P76}. Targeted data augmentation (e.g., SMOTE) is also used to populate parts of the data distribution that are underrepresented \cite{P77}.
\subsection{Data Analysis}
In the data analysis phase, the results are interpreted and knowledge is extracted. In the following we describe the biases that occur in this phase.
\subsubsection{Deployment Bias}
Deployment Bias occurs after model deployment, when a system is used or interpreted in inappropriate ways. A model is built to carry out a particular task but what happens when an autonomous model is moderated by institutional structure (also called Framing Trap) \cite{P78} or its recommendations would be interpreted erroneously by humans, a deployment bias will occur \cite{P79}.
\paragraph{Example:}Risk assessment tools in the criminal justice system predict a risk score, but a judge may interpret this in unexpected ways before making his or her final decision \cite{P15}.
\paragraph{Best practice:}Testing the model in the real world environment can aid to minimize its harms. User training of the model could be an effective step to handle deployment bias. Ethical model training using unbiased and transparent data and careful planning can be helpful to get optimal results \cite{P80}.
\subsubsection{Rescue Bias}
Rescue bias is an interpretive bias related to analyzer's personal preconceptions that pumps him to discount the data by finding selective faults in a trial that undermines his/her expectations. In other words, it is a planned attempt to emasculate the findings and draw the pre-planned conclusions. One may fall for rescue bias after seeing the unexpected or below par ends of the experimentation \cite{P85}.
\paragraph{Example:}Binge eating disorder could be excluded from ICD-11 due to arguments that it might stigmatize people who eat a lot or individuals who have a high body mass index. However, given the elevated mortality and other health risks associated with eating disorders, this would have a significant adverse impact, particularly on young women \cite{P86}.
\paragraph{Best practice:}To control the behavioral biases like this, data experts should think critically and try to be logical. Self-criticism or self-evaluation is one of the efficient procedures to not be biased \cite{P45}.
\subsubsection{Overfitting and Underfitting}
Overfitting and Under-fitting are two problems with machine learning models. Overfitting is when the model grabs the trend of training data patterns so well that it does not improve its ability to solve problem anymore. Under-fitting is reverse of overfitting. It is the statistical model that cannot capture the underlying trend of data or model over idealizes its experience \cite{P81}.
\paragraph{Example:}Lets assume we want to predict if a student will land a job interview based on her resume. Then, we assume that we train a model from a dataset of 10000 resumés and their outcomes. Then, we try the model out on the original dataset, and it predicts outcomes with 99 percent accuracy. When we run the model on a new (\textit{unseen}) dataset of resumes, we only get 50 percent accuracy \cite{P82}.
\paragraph{Best practice:}Process Mining Algorithms can reduce the gap between overfitting and under-fitting \cite{P83}. In \cite{P84}, authors presented two important techniques namely Penalty Methods, and Early Stopping, to limit these effect (overfitting and under-fitting). Furthermore, a novel L1/4 regularization method to overcome this issue, is stated in \cite{P81}.

\section{Future Work}
In this research paper we analyzed different phases of the data science lifecycle. Which biases can occur
in these phases? We have provided all these biases with a description, examples and best practices.
In the further research we are dealing further with mitigation methods. How can data scientists avoid all
biases in this document?
Furthermore, we create a vocabulary from the findings of these researchers to create an knowledge
graph for data scientists. This graph should help them to identify relations between each keywords.

\bibliographystyle{plain}%
\bibliography{Literatur}

\end{document}